\documentclass{elsart}
\usepackage{natbib}
\usepackage{epsfig}
\begin{document}
\runauthor{Cicero, Caesar and Vergil}
\begin{frontmatter}
\title{
An investigation of triply heavy baryon production at hadron
colliders}
\author{M.A. Gomshi Nobary$^{a,b}$ and R. Sepahvand$^{a}$}
\thanks[]{E-mail: mnobary@razi.ac.ir and mnobary@aeoi.org.ir}
\address{$^a$ Department of Physics, Faculty of Science, Razi University, Kermanshah, Iran.}
\address{$^b$ The Center for Theoretical Physics and Mathematics, A.E.O.I.,
Roosbeh Building, P.O. Box 11365-8486 Tehran, Iran. }
\begin{abstract}
The triply heavy baryons have a rather diverse mass range. While
some of them possess considerable production rates at existing
facilities, others need to be produced at future high energy
colliders. Here we study the direct fragmentation production of
the
 $\Omega_{ccc}$ and
$\Omega_{bbb}$ baryons as the prototypes of triply heavy baryons
at the hadron colliders with different $\sqrt{s}$. We present and
compare the transverse momentum distributions of the differential
cross sections, $p_T^{\rm min}$ distributions of total cross
sections and the integrated total cross sections of these states
at the RHIC, the Tevatron Run II and the CERN LHC.
\end{abstract}
\end{frontmatter}

\section{Introduction}
Study of hadron production and decay has always been interesting.
Historically it has served to illuminate both the collider physics
and the fundamental theories of interactions specially in the
strong and electroweak sectors. Recently heavy hadrons have
received great attention. Their structure is predicted by the
constituent quark model and, wherever light quarks are absent,
they are nicely treated within the framework of the effective
field theory and the perturbative QCD [1].

Singly and doubly heavy baryons, $\Lambda$'s and $\Xi$'s , exhibit
interesting properties and due to the involvement of the light
degrees of freedom, they have been studied in special models [2].
Triply heavy baryons are the heaviest composite states predicted
by the constituent quark model. They are the last generation of
baryons within the standard model. Essentially they are the
$\Omega_{ccc}$, $\Omega_{ccb}$, $\Omega_{cbb}$ and $\Omega_{bbb}$
baryons. It has become clear that while the $\Omega_{ccc}$ and
$\Omega_{bbb}$ should be fragments of a $c$ and a $b$ quark
respectively, the $\Omega_{ccb}$ and $\Omega_{cbb}$ may be
produced each in a $c$ or in a $b$ quark fragmentation. The
fragmentation of all triply heavy baryons has recently been
studied and their production has been estimated at the CERN LHC
[3]. However the wide range of their fragmentation probabilities
($10^{-4}$ - $10^{-7}$), higher cross section of charm production
at the Tevatron and RHIC and finally different acceptance cuts
apart from different $\sqrt{s}$ for different colliders, suggest
their production to be explored in other colliders as well. For
the matter of simplicity, in this work we have chosen to
investigate the production of the lightest ($\Omega_{ccc}$) and
the heaviest ($\Omega_{bbb}$) as prototypes of the triply heavy
baryons at the RHIC, the Tevatron Run II and the CERN LHC
colliders.

\section{Fragmentation functions}
To evaluate the cross section of $\Omega_{ccc}$ and $\Omega_{bbb}$
baryons at hadron colliders in a factorized scheme, we need their
fragmentation functions. Indeed, it is possible to describe the
fragmentation of these states by a single fragmentation function
for $Q\rightarrow\Omega_{QQQ}$ which has already been calculated
[4]. In our model we have considered the emission of two gluons by
the heavy quark $Q$, each producing a $\overline Q Q$ pair. The
three heavy quarks thus obtained form the $\Omega_{QQQ}$ bound
state leaving the heavy anti-quarks to form the final state jet.
The bound state is characterized by the baryon decay constant
$f_B$. The fragmentation process in leading order is described in
Figure 1. Kinematically we have let the original heavy quark to
keep its original transverse momentum $q_T$ and have ignored the
respective motion of the constituents within the bound state. This
perturbative picture is evaluated at the scale $\mu =\mu_\circ $.
The scale $\mu_\circ$ which is a scale at which such calculations
are possible, is in the order of total mass of all final state
particles, namely $5m_c$ for $c\rightarrow\Omega_{ccc}$ and $5m_b$
for $b\rightarrow\Omega_{bbb}$ states. The result is the following
fragmentation function

\begin{figure}
\begin{center}
\includegraphics[width=14cm]{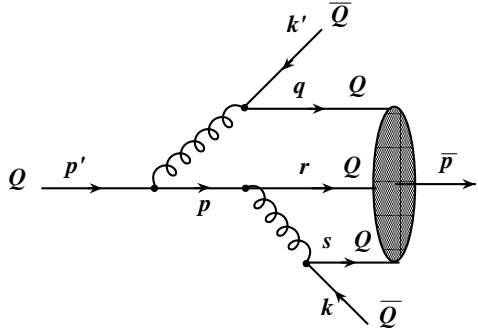}
\caption[Submanifold]{The lowest order Feynman diagram
illustrating the fragmentation process of a heavy quark $Q$ into a
triply heavy baryon $\Omega_{QQQ}$ with identical constituent
flavors. } \label{f.emb}
\end{center}
\end{figure}

\begin{eqnarray}
D_{Q\rightarrow
QQQ}(z,\mu_\circ)&=&{{\pi^4\alpha_s^4(2m_Q)f_B^2C_F^2}\over {108
m^2z^4(1-z)^4 f(z)^2 g(z)^6}}\nonumber\\
&&\times\bigl[\xi^8 z^8+4\xi^6 z^6(83-130 z+51 z^2)\nonumber\\
&&+6\xi^4z^4(1413-3084z+3022z^2-2156z^3+821z^4)\nonumber\\
&&+4\xi^2 z^2(18711-51678z+69417z^2-70308z^3\nonumber\\
&&+53529z^4-25950z^5+6343z^6)+222345-740664z\nonumber\\
&&+1179036z^2-1253448z^3+90126z^4-388872z^5\nonumber\\
&&+109916z^6-49912z^7+20649z^8\bigr],
\end{eqnarray}

\noindent where $\alpha _s$ is the strong interaction coupling
constant evaluated at the pair of vertices of each gluon in the
Figure 1, $f_B$ is the baryon decay constant which is defined in a
similar manner to the meson decay constant, $f_M$ and $C_F$ is the
color factor of the baryon state formed in the fragmentation of
the heavy quark. Moreover, here we have defined $\xi=\langle{
q}_T^2\rangle/m^2$ with $q_T$ being the transverse momentum of the
initial heavy quark and $m$ is the heavy quark mass. The two
functions $f(z)$ and $g(z)$ are defined as

\begin{eqnarray}
f(z)= \frac{-\langle q_T^2\rangle}{{3m^2}}+ {3\over z}+
{{4}\over{3}} \biggl[1+\frac{\langle q_T^2\rangle}{
4m^2}\biggr]{1\over{1-z}},\quad g(z)=-\frac{1}{3}+f(z).
\end{eqnarray}

The function $f(z)$ is the contribution of the energy denominator
emerging from the phase space integration and the function $g(z)$
is due to the quark and gluon propagators.

The inputs for the fragmentation function (1) are the quark mass,
baryon decay constant and the color factor. We have set
$m=m_c=1.25$ GeV and $m=m_b=4.25 $ GeV. For the decay constant and
the color factor we have taken $f_B$=0.25 GeV and $C_F=7/6$ for
both cases of the $\Omega_{ccc}$ and $\Omega_{bbb}$ states.

\section{Inclusive production of a $\Omega_{QQQ}$ baryon}
Inclusive production of a $\Omega_{QQQ}$ baryon state in a
hadronic collision is fulfilled in certain stages. Collision of
hadrons provide the production of the required parton or partons
which eventually fragment into a $\Omega_{QQQ}$ state. The
theoretical evaluation of this process is only possible at
sufficiently high transverse momentum and taking advantage of the
parton model factorization. All this is possible at a scale much
higher than the calculable scale of the fragmentation functions.
Therefore the fragmentation functions which are calculated at
fragmentation scale, are evolved up to a scale at which the
convolution of parton distribution functions, bare cross section
of the initiating heavy quark, and the fragmentation function is
possible. In other words, for the $pp$ collision we may write

\begin{eqnarray}
\frac {d\sigma}{dp_T}\bigl[p p &\rightarrow &\Omega_{QQQ}(p_T)+
X\bigr]=\sum_{i,j}\int dx_i dx_j dz f_{i/p}(x_i,\mu)f_{j/
p}(x_j,\mu)\nonumber\\ &&\times\bigl[ \hat\sigma(ij\rightarrow
Q(p_T/z)+X,\mu) D_{ Q\rightarrow \Omega_{QQQ}}(z,\mu)\bigr].
\end{eqnarray}

Here $f_{i/p}(x_i,\mu)$ and $f_{j/ p}(x_j,\mu)$ are the parton
distribution functions for the initial partons $i$ and $j$
carrying fractions $x_i$ and $x_j$ of the total momentum in the
protons, $\hat\sigma$ is the heavy quark production cross section
and $D_{ Q\rightarrow \Omega_{QQQ}}(z,\mu)$ represents the
fragmentation of the produced heavy quark into a triply heavy
baryon. Note that here the scale $\mu$ in the parton distribution
functions, subprocess cross sections and the fragmentation
functions are set to be equal. For the parton distribution
functions we have employed the parameterization due to
Martin-Roberts-Stirling (MRS) [5],  and have included the heavy
quark production cross section up to the order of $\alpha_s ^3$
[6]. The $\sqrt{s}$ and acceptance cuts for the colliding
facilities used in this work appear in Table 1.

The production rates should not depend on the choice of the scale
$\mu$ if both the production of high-energy partons and the
fragmentation functions in all orders in $\alpha_s$ in the
perturbation expansion are included. However, only the results for
next leading order parton production cross section and the leading
order fragmentation function are available. Therefore the
 results obtained form (3) will depend on the scale $\mu$. It is usual
to choose the transverse mass of the heavy quark as the central
choice of scale defined by $\mu_R=\sqrt{ {p_T}^2{\rm
(parton)}+{m_Q}^2 }$. If such a choice or sometimes multiple of it
happens to be less than the fragmentation scale, then the larger
of ($\mu,\;\mu_\circ$) is chosen as appropriate scale. We have
used the following form of the Altarelli-Parisi equation [7] to
evolve our fragmentation functions

\begin{eqnarray}
\mu \frac{\partial}{\partial \mu} D_{Q \rightarrow QQQ} (z,\mu)=
 \int_z^1 \frac{dy}{y} P_{Q \rightarrow Q}
(z/y,\mu) D_{Q \rightarrow QQQ}(y,\mu).
\end{eqnarray}
\noindent Here $P_{Q\rightarrow Q}(x=z/y,\mu) $ is the
Altarelli-Parisi splitting function. Note that only the term $P_{Q
\rightarrow Q}$ is included in (4). The reason is that the quark $
Q$ is assumed to be heavy enough to make other contributions
irrelevant. The boundary condition on the evolution equation (4)
is the initial fragmentation function $D_{Q \rightarrow
QQQ}(z,\mu)$ evaluated at the fragmentation scale $\mu=\mu_\circ$.
It should also be mentioned that the evolution of the
fragmentation functions also sums up the logarithms of the order
of $\mu_R/m_Q$ in the fragmentation functions.

To check the sensitivity of our results, we have examined the
behavior of the differential cross sections of $\Omega_{ccc}$ and
$\Omega_{bbb}$ at the RHIC, the Tevatron Run II and the LHC for
different scales ranging from the fragmentation scale up to the
scales below the $Z^\circ$ boson mass. It is seen that the
sensitivity decreases with increasing scale. Such a study have led
us to select appropriate scales for our further investigations. In
this way we were motivated to choose the scales of $4\mu_R$ and
$6\mu_R$ for the $\Omega_{ccc}$ and $\Omega_{bbb}$ states
respectively.

\section{Results and discussion}
In summary we have considered $\Omega_{ccc}$ (the lightest) and
$\Omega_{bbb}$ (the heaviest) triply heavy baryons to evaluate
their production rates at the hadron colliders. To accomplish
this, we have employed the next leading order results for parton
production cross sections. The fragmentation functions used here
are calculated in leading order perturbation theory. They provide
reliable fragmentation probabilities for the triply heavy baryons.
To evaluate the cross sections, the well known patron model
factorization at high transverse momentum is employed. This
procedure allows the complicated mechanism of hadron production to
be treated in a factorized manner. This is possible in a scale
which is higher than the scale at which the fragmentation
functions are calculated. The Altarelli-Parisi evolution equation
relates these scales. In the evolution of the fragmentation
functions we have included only the $P_{Q\rightarrow Q}$ splitting
function. Here, our evaluation of cross sections for the triply
heavy baryons is devoted to different hadron colliders. Each
collider with detection system has restrictions on the
measurements of the transverse momentum and the rapidity of the
particles. The so called acceptance cuts for the colliders
considered here appear in Table1.

First we present the transverse momentum, $p_T$, distributions of
the differential cross sections at different hadron colliders for
$\Omega_{ccc}$ and $\Omega_{bbb}$. They appear in Figures 2 and 3.
Clearly, the distributions are sensitive to the choice of $\mu$.
Our choice of $\mu=4\mu_R$ for $\Omega_{ccc}$ and $\mu=6\mu_R$ for
$\Omega_{bbb}$ is at the region of the scale with minimum
sensitivity. Although the cross section for a given $p_T$ differs
up to three orders of magnitude from one collider to the other,
 they are still comparable. The difference is seen
to grow with increasing $p_T$. It is more interesting in the case
of $\Omega_{ccc}$ where the distributions seem to converge at
sufficiently low $p_T$. Another important feature about these
distributions is the rather high cross section of $\Omega_{ccc}$
which is more striking in the case of RHIC where low $p_T$'s are
available. Figures 4 and 5 show the total cross sections for
production of $\Omega_{ccc}$ and $\Omega_{bbb}$ with transverse
momentum above a minimum value $p_T^{\rm min}$. Note that in
Figures 2 and 3 only the range $p_T>p_T^{\rm min}=p_T^{\rm cut}$
were considered. It seems that the above mentioned general
features, hold in the case of $p_T^{\rm min}$ distributions,
unless for that the difference in the differential cross sections
for a given $p_T$ is not as much as that of  $p_T^{\rm min}$
distributions. Another general feature of our distributions is
that the fall off decreases with increasing $\sqrt{s}$ and also
with increasing $p_T^{\rm min}$ or $p_T$.

\begin{table}
\caption{The center of momentum energy ($\sqrt{s}$) and the
acceptance cuts for the colliding facilities used in this work.
The rapidity is defined as $y=\frac{1}{2}
\log\bigl\{({E-p_L})/({E+p_L})\bigr\}$. }
\begin{center}
\begin{tabular}{l l l l}
\hline
   &RHIC& Tevatron Run II& CERN LHC\\
\hline $\sqrt{s}$ [GeV]&$\;$200&\qquad1960&$\;$\quad14000\\
 $p_T^{\rm cut}$[GeV] & $\;\;\; 2$ & $\quad\;\;\; \quad 6$& $\qquad\; 10$\\
$\quad\;\; y\leq$ & $\;\;\; $3  & $\quad\;\;\; $\quad 1&$ \qquad\;\; 1$\\
\hline
\end{tabular}
\end{center}
\end{table}

\begin{figure}
\begin{center}
\includegraphics[width=12cm]{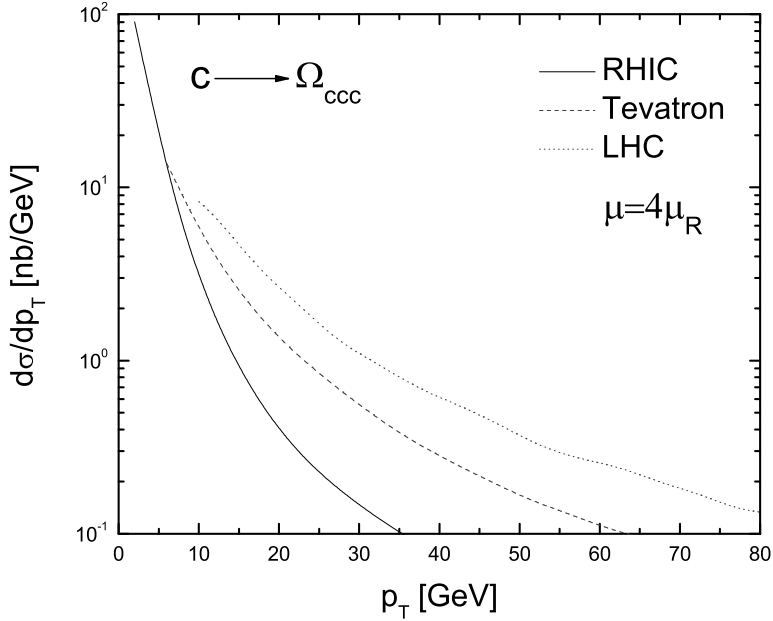}
\caption[Submanifold]{The $p_T$ distributions of the differential
cross sections in nb for $\Omega_{ccc}$ production at the RHIC,
the Tevatron Run II and the CERN LHC hadron colliders at the scale
of $\mu=4\mu_R$. } \label{f.emb}
\end{center}
\end{figure}

\begin{figure}
\begin{center}
\includegraphics[width=12cm]{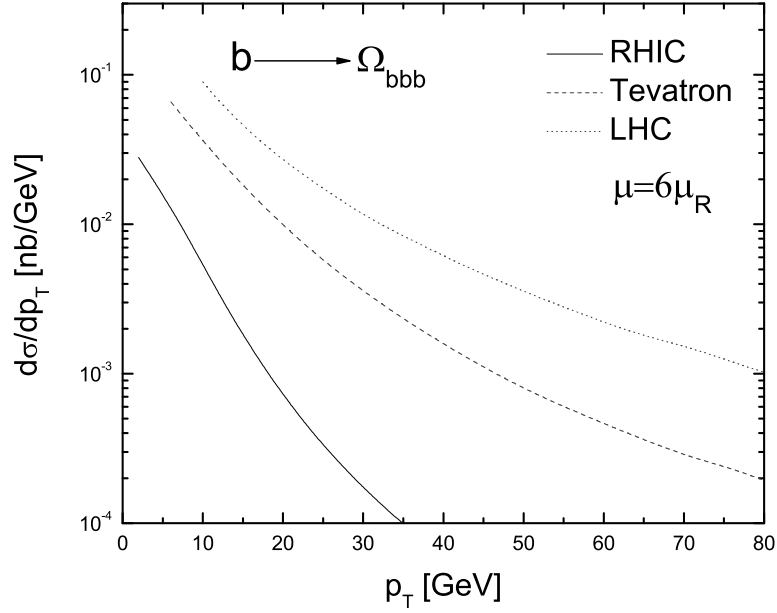}
\caption[Submanifold]{The same as Fig.2 but in the case of
$\Omega_{bbb}$ production at the scale of 6$\mu_R$. }
\label{f.emb}
\end{center}
\end{figure}

\begin{figure}
\begin{center}
\includegraphics[width=12cm]{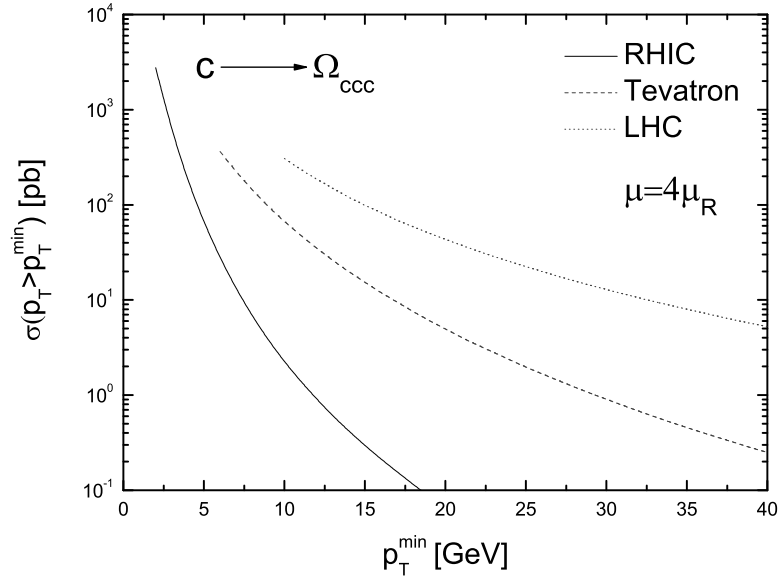}
\caption[Submanifold]{The $p_T^{\rm min}$ distribution of the
total cross section in pb for $\Omega_{ccc}$ production at the
RHIC, the Tevatron Run II and the CERN LHC hadron colliders at the
scale of $\mu=4\mu_R$. } \label{f.emb}
\end{center}
\end{figure}

\begin{figure}
\begin{center}
\includegraphics[width=12cm]{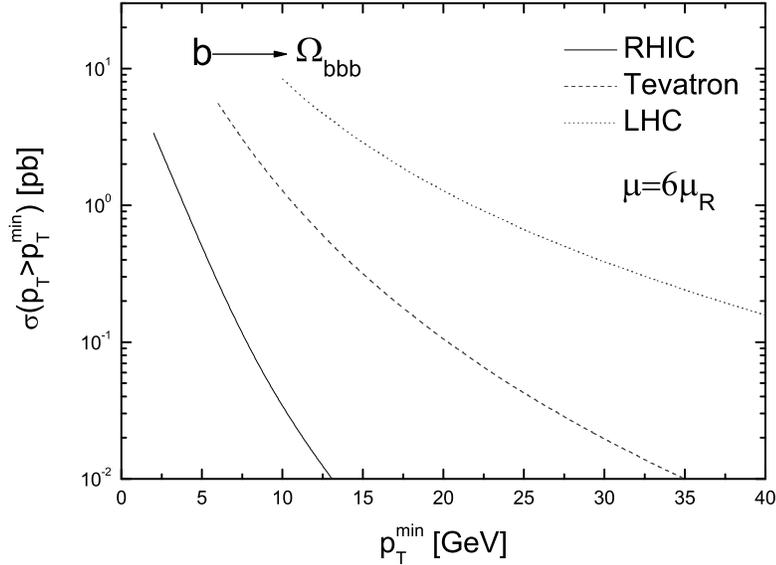}
\caption[Submanifold]{The same as Fig. 4 but for the case of
$\Omega_{bbb}$ at the scale of $\mu=6\mu_R$. } \label{f.emb}
\end{center}
\end{figure}

\begin{table}
\caption{The total integrated cross section in pb for $
\Omega_{ccc}$ and $\Omega_{bbb}$ baryons at different hadron
colliders. The decimal places may not be significant within the
uncertainties involved. } \vskip .5cm
\begin{center}
\begin{tabular}{l l l l}
\hline
 {\rm Process}  &$\;\;$RHIC& Tevatron Run II& CERN LHC\\
\hline
 $c\rightarrow \Omega_{ccc}\;({\rm at}\mu=4\mu_R)$ &$\;$ 2758.3 &$\;\;\;\;$ 383.0& $\;\;$ 308.0\\
$b\rightarrow \Omega_{bbb}\; ({\rm at}\mu=6\mu_R)$ & $\;\;\;\;\;$3.3 & $\;\;\;\;\;\;\;$6.0&$\;\;\;\;$ 8.4\\
\hline
\end{tabular}
\end{center}
\end{table}

We have also calculated the integrated total cross sections. They
appear in Table 2. Note that the cross section for $\Omega_{bbb}$
increases with increasing $\sqrt{s}$. But this is not the case for
$\Omega_{ccc}$. Not only the order reverses in this case, but the
cross section at RHIC is nearly one order of magnitude higher.

What can we say about other triply heavy baryons? It seems that
the baryons with at least two $c$ or at least two $b$ quarks,
apart from the magnitude of their cross sections, behave in a
similar fashion [3]. Therefore we expect that the $\Omega_{ccb}$
state produced in $c$ or $b$ quark fragmentation to have similar
distributions as $\Omega_{ccc}$. Likewise the $\Omega_{cbb}$ state
emerging from a $c$ or a $b$ quark will behave like $\Omega_{bbb}$
state. It is also interesting to note that our choice of
$\Omega_{ccc}$ and $\Omega_{bbb}$ with fragmentation probabilities
of about $2\times 10^{-5}$ and $6\times 10^{-7}$ are not the
states with maximum and minimum fragmentation probabilities. In
other words, the lightest/heaviest of triply heavy baryons does
not mean the state with maximum/minimum fragmentation
probabilities. Indeed the fragmentation probabilities for
$b\rightarrow\Omega_{ccb}$ and $c\rightarrow\Omega_{cbb}$ possess
the maximum and the minimum fragmentation probabilities of
$2\times 10^{-4}$ and $10^{-7}$ respectively.

At the end we would like to discuss the reliability of our
results. We have employed a fragmentation function obtained in a
particular model. Without referring to its ingredients, it is
worth mentioning that the predictions of this model is consistent
with similar results in the case of doubly heavy baryons evaluated
by Doncheski et al in [2] and also experimental results of [9].
This kind of comparison is presented in [3]. Moreover, similar
fragmentation functions predict consistent fragmentation
probabilities for $J/\psi$ and $\Upsilon$ states [8]. The
remaining uncertainties are related to the calculation of the
cross sections. Here the the parton densities and the scales are
relevant. These kinds of uncertainties are well described in the
literature. Therefore, we believe the only factors which may alter
our results are the consideration of higher order fragmentation
functions and the heavy quark production cross sections.

\end{document}